# NEURAL NETWORKS FOR ANATOMICAL THERAPEUTIC CHEMICAL (ATC) CLASSIFICATION

Loris Nanni[1], Alessandra Lumini[2], and Sheryl Brahnam[3*]

[1]DEI at the University of Padova, Via Gradenigo, 6, Padova, Italy 35131. E-mail: loris.nanni@unipd.it.

[2]DISI at the University of Bologna, Via dell'Università 50, Cesena, Italy 47521. E-mail: alessandra.lumini@unibo.it.

[3]ITC Department, Glass Hall Rm 387, 901 S National, Missouri State University, Springfield, MO 65801. E-mail: sbrahnam@missouristate.edu.

**Abstract**

Objective: The aim of this paper is to experimentally derive an ensemble of different feature descriptors and classifiers for Automatic Anatomical Therapeutic Chemical (ATC) classification that outperforms the state-of-the-art.

Impact: We succeed in our objective by developing an ensemble for ATC classification that is composed of many multi-label classifiers trained on distinct sets of ATC descriptors, including features extracted from a Bidirectional Long Short-Term Memory Network (BiLSTM). The MATLAB source code of our system is freely available to the public.

Introduction: Automatic Anatomical Therapeutic Chemical (ATC) classification is progressing at a rapid pace because of its potential in drug development. Predicting an unknown compound's therapeutic and chemical characteristics according to how they affect different organs and physiological systems makes automatic ATC classification a vital but challenging multi-label problem.

Methods: Ensembles are generated and compared that combine multi-label classifiers based on Multiple Linear Regression (hML) with LSTM classification and features taken from LSTM. both classifiers are trained on DDI, FRAKEL, and NRAKEL descriptors. Features extracted from the trained LSTM are also fed into hML. Evaluations are tested on a benchmark data set of 3883 ATC-coded pharmaceuticals taken from KEGG, a publicly available drug databank.

Results: Experiments demonstrate the power of our best ensemble, which is shown to outperform the best methods reported in the literature, including the state-of-the-art developed by the fast.ai research group.

Conclusion: This study demonstrates the power of extracting LSTM features and combining them with ATC descriptors in ensembles for ATC classification.

**Keywords:** Bioinformatics, Machine learning, Multi-label classifier, Bidirectional long short-term memory, ATC classification, Learned Features.

# 1 Introduction

From start to market, the price for engineering new drugs, which can take decades before final approval, is now estimated to be 2.8 billion USD [1]. Of all drugs currently under development, approximately 86% will fail to be better than placebo [2] or will prove to cause more harm than good [3]. To weed out new drugs with a low probability of being efficacious and safe has led researchers to search for automatic methods for classifying compounds according to the organs they are likely to affect based on these compounds' Anatomical Therapeutic Chemical (ATC) classes. An automatic classification system with good ATC prediction would not only accelerate research but also significantly reduce drug development costs.

The ATC coding system [4] classifies compounds into one or more classes at five levels in terms of the drug's effects on organs or physiological systems. Most relevant to the automatic ATC classification problem is the first ATC level, which determines the general anatomical groups, as coded with fourteen semi-mnemonic letters that a particular compound targets. These alphabetic codes range from **A** (alimentary tract and metabolism) to **V**, a category that includes various groups. Levels 2/3 are pharmacological subgroups, and levels 4/5 contain chemical subgroups. A compound is assigned to as many ATC codes as relevant within each of these five levels.

Despite the serviceability of the ATC classification system for assessing the clinical value of a compound, most pharmaceuticals have yet to be assigned ATC codes. Accurate coding involves expensive, labor-intensive experimental procedures. Hence, the pressing need for machine learning (ML) to be applied to this problem.

Early ML systems tended to simplify the complexity of the ATC classification problem by reducing the level 1 multi-class problem to a single class problem. Dunkel et al. [5], for example, took advantage of a compound's unique structure to identify its class, while Wu et al. [6] based their approach on extracting relationships among level 1 subgroups. Chen et al. [7] tackled the multi-label complexity of ATC classification by examining a drug's chemical-chemical interactions. The authors also established the de facto benchmark data set for ATC classification. Cheng et al., in [8] and [9], designed ML systems to handle class overlapping by fusing different descriptors: structural similarity, fingerprint similarity, and chemical-chemical interaction. Nanni and Brahnam [10] transformed these same 1D vectors into images (matrices) and extracted texture descriptors from them. The descriptors were then trained on ensembles of multi-label classifiers.

Convolutional Neural Networks (CNNs) were trained on 2D descriptors in [11] and in [12], but in Lumini and Nanni [12], a set of features were extracted from deep leaners for training two multi-label classifiers. This approach was further expanded in Nanni, Brahnam, and Lumini [13]. Ensembles of CNNs were constructed by adjusting batch sizes and learning rates, and different methods were applied to handle multi-label inputs.

In this work, an ensemble of different feature descriptors and classifiers is proposed that strongly outperforms the state-of-the-art classification results on the ATC benchmark data set developed by Chen et al. [7]. The system proposed here was experimentally developed by comparing and evaluating multi-label classifiers trained on different feature/descriptor sets. Our best results were obtained by combining a Bidirectional Long Short-Term Memory Network (BiLSTM) [14] with a multi-label classifier.

## 2 Methods

The approach taken in this study is to produce experimentally ensembles that combine multi-label classifiers (hML) based on Multiple Linear Regression with LSTM classification and features taken from LSTM and trained on hML. All these classifiers are trained on $X$, a set of three different descriptros (DDI, FRAKEL, NRAKEL, detailed in section 3.1). The features extracted from LSTM are also fed into hML classifiers. As illustrated in Figure 1, the results are then combined and evaluated.

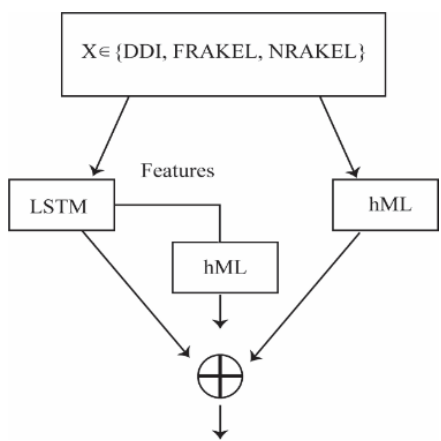

Figure 1. Schematic of proposed ATC classification approach.

The LSTM feature extraction process and multi-label classifiers are discussed in sections 2.1-2.2. In this work, we also examine the FastAI Tabular model [15], detailed in section 2.3, a method which has obtained the best classification result on the Chen benchmark [7].

*2.1 LSTM multi-label classifier and feature extractor*

LSTM is a Recurrent Neural Network that makes a decision for what to remember at every time step. As illustrated in Figure 2, this network contains three gates: 1) input gate $I$, 2) output gate $O$, and 3) forget gate $f$, each of which consist of one layer with the sigmoid ($\sigma$) activation function. LSTM also contains a specialized single layer network candidate $\bar{C}$, which has a $Tanh$ activation function. In addition, there are four state vectors: 1) memory state $C$ with 2) its previous memory state $C_{t-1}$ and 3) hidden state $H$ with 4) its previous state $H_{t-1}$. The varable $X$ in Figure 2 represents the current input at time step $t$.

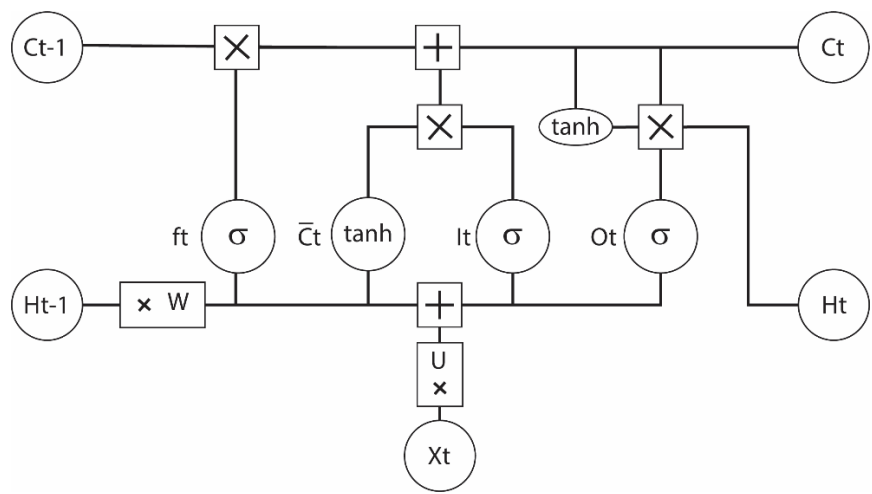

Figure. 2. Long Short-Term Memory (LSTM) classifier.

The process for updating LSTM at time $t$ is as follows. Given $X_t$ and $H_{t-1}$ and letting $U, W, b$ be the learnable weights of the network (each independent of $t$), the candidate layer $\bar{C}_t$ is

$$\bar{C}_t = Tanh(U_c X_t + W_c H_{t-1} + b_c). \tag{1}$$

The next memory cell is

$$C_t = f_t * C_{t-1} + I_t * \bar{C}_t, \tag{2}$$

where ∗ is element-wise multiplication.

The gates are defined as

$$f_t = \sigma(U_f X_t + W_f H_{t-1} + b_f), \qquad (3)$$

$$I_t = \sigma(U_i X_t + W_i H_{t-1} + b_i); \qquad (4)$$

$$O_t = \sigma(U_o X_t + W_o H_{t-1} + b_o). \qquad (5)$$

The output is $H_t = O_t * \sigma(C_t)$ of $O_t$ and the sigmoid of $C_t$.

Regarding input, all sequences for this task are of the same length, so sorting input by length is not required. The output of LSTM can be the entire sequence $H_t$ (this permits several layers to be stacked in a single network) or the last term of this sequence.

An LSTM that has two stacked layers trained on the same set of samples is called a Bidirectional LSTM (BiLSTM). The second LSTM connects to the end of the first sequence and runs in reverse. BiLSTM is best used to train data not related to time. Accordingly, this study uses the BiLSTM, as implemented in the MATLAB LSTM toolbox. Parameters were set to the following values: $numHiddenUnits = 100$, $numClasses = 14$, and $miniBatchSize = 27$.

LSTM is not ordinarily considered a multi-label classifier but can perform multi-label classification if the training strategy outlined in [13] is implemented, which involves replicating a sample $m$ times for each of its $m$ labels. To assign a test pattern to more than one class, a rule is applied in the final softmax layer where a given pattern is assigned to each of the classes whose score is larger than a given threshold.

LSTM can function not only as a classifier but also as a feature extractor. As noted in Figure 1, in this study LSTM functions in both capacities. Feature extraction with LSTM is accomplished by representing each pattern using the activations from the last layer, which produce a feature vector with a dimension equal to the number of classes. Feature perturbation and extraction are performed several times by randomly sorting the original set of features used to train the LSTM.

*2.2 Classification by hML*

The algorithm hML-KNN, proposed in [16], is a multi-label classifier that integrates a feature score and a neighbor score. The feature score decides if a sample belongs to a particular class using the global information contained in the whole training set. In contrast, the neighbor score decides a sample's class labels based on the class assignment of its neighbors. The feature score $f_1(x, g_j)$ for a given pattern $x$ with respect to an anatomical group $g_j$ is calculated to evaluate whether the pattern belongs to the group $g_j$ using a regression model. The neighbor score $f_2(x, g_j)$ calculates the significance of the class membership of K neighbors of a pattern belonging to a given group $g_j$: the neighbor score increases if more neighbors of $x$ have the label $g_j$. Thus, $f_2(x, g_j)$ is 1 if all neighbors of $x$ belong to $g_j$, 0 otherwise. The final score of $x$ is a weighted sum of the two factors:

$$f(x, gj) = \alpha f_1(x, g_j) + (1 - \alpha) f_2(x, g_j) \qquad (6)$$

In our experiments, we use the default values where the weight factor $\alpha$ is set to 0.5, and the number of neighbors is K=15.

### 2.3   Classification by FastAI Tabular Model

In addition to hML and LSTM, we explore the FastAI Tabular model [15], which is a powerful deep learning technique for tabular/structured data based on the creation of some embedding layers for categorical variables. This deep learner uses embedding layers to represent categorical variables by a numerical vector whose values are learned during training. Embeddings allow for relationships between categories to be captured, and they can also serve as inputs to other models.

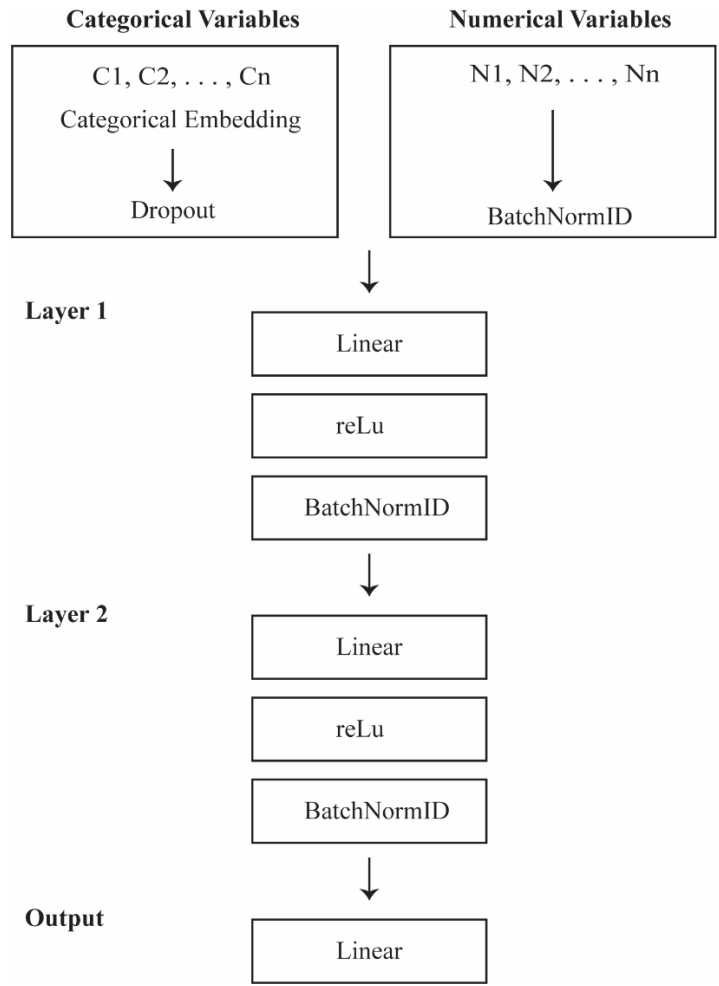

Figure 3. Schematic of the FastaAI Tabular model

A graphic representation of a FastAI Tabular model is presented in Figure 3, where one can observe that the categorical variables are transformed into N-dimensional features by categorical embeddings followed by a dropout layer to prevent overfitting. Numerical variables are simply normalized. Then all the variables are concatenated and passed as input into the following layers, which, in our experiments, are two hidden layers and one output layer, as illustrated in Figure 3. We also use a binary encoding to represent binary variables, and the resulting variable is treated as categorical.

## 3 Experimental results

3.1 *The data set and descriptors*

The ensembles generated by the proposed approach are compared and evaluated on the data set in [7] (Supporting Information S1). This data set is a collection of 3883 ATC-coded pharmaceuticals taken from KEGG [17], a publicly available drug databank. The most classes any one compound belongs to is six. The total number of drugs with more than one label is 4912. This virtual subset based on the number of samples is called N(Vir). The average number of labels per sample is thus 4912/3883=1.27.

The following descriptors represent the drugs in this data set:

- **DDI** represents each drug with three mathematical expressions representing the maximum interaction score with the drugs, the maximum structural similarity score, and the molecular fingerprint similarity score, with each expression based on its correlation with the 14 level 1 classes. Thus, the resulting descriptor is of size 14×3=42 (available in the supplementary material in Nanni and Brahnam [10]).
- **FRAKEL** represents each drug by its ECFP fingerprint [18], which is a 1024-dimensional binary vector (located at http://cie.shmtu.edu.cn/iatc/index). The descriptor is obtained by feeding the drug into RDKit (http://www.rdkit. org/), a free ML toolkit for chemistry informatics. From this 1024-dimensional binary vector, a 64- dimensional categorical descriptor is obtained, representing each group in 16 bits as an integer. This version of FRAKEL has been used with the FastAI Tabular model.
- **NRAKEL** represents a drug by a 700-dimensional descriptor obtained from the Mashup algorithm [19], which generates output from seven drug networks (five based on chemical-chemical interaction and two on drug similarities).

3.2 *Testing protocol*

The jackknife testing protocol is used here to generate both the training and testing sets. At each iteration of this protocol, one sample is placed in the testing set and the remainder in the training set. Iteration continues until each pattern has taken a turn in the testing set. The K-fold cross-validation is also applied. The jackknife protocol was selected as stipulated in [20].

3.3 *Performance indicators*

ATC classification is evaluated using the standard performance indicators defined in [20] and repeated below:

$$\text{Aiming} = \frac{1}{N}\sum_{k=1}^{N}\left(\frac{\|\mathbb{L}_k \cap \mathbb{L}_k^*\|}{\|\mathbb{L}_k^*\|}\right), \quad (7)$$

$$\text{Coverage} = \frac{1}{N}\sum_{k=1}^{N}\left(\frac{\|\mathbb{L}_k \cap \mathbb{L}_k^*\|}{\|\mathbb{L}_k\|}\right), \quad (8)$$

$$\text{Accuracy} = \frac{1}{N}\sum_{k=1}^{N}\left(\frac{\|\mathbb{L}_k \cap \mathbb{L}_k^*\|}{\|\mathbb{L}_k \cup \mathbb{L}_k^*\|}\right), \quad (9)$$

$$\text{Absolute True} = \frac{1}{N}\sum_{k=1}^{N}\Delta(\mathbb{L}_k, \mathbb{L}_k^*), \quad (10)$$

$$\text{Absolute False} = \frac{1}{N}\sum_{k=1}^{N}\left(\frac{\|\mathbb{L}_k \cup \mathbb{L}_k^*\| - \|\mathbb{L}_k \cap \mathbb{L}_k^*\|}{M}\right), \quad (11)$$

where $M$ the number of classes, $N$ is the number of samples, $\mathbb{L}_k$ is the true label, $\mathbb{L}_k^*$ is the predicted label, and $\Delta(\cdot,\cdot)$ returns 1 if the two sets have the same elements, 0 otherwise.

3.4 *Experiments*

The first experiment (see Table 1) compares the three multi-label classifiers described in section 2. Also compared are three other standard classifiers, each trained on the three sets of features (DDI, FRAKEL, and NRAKEL). As already mentioned, LSTM is not a native multi-label classifier; thresholding was used as described in section 2.1 to adapt this classifier to the ATC classification problem.

The six classifiers reported in Table 1 are the following:

- RR, a Ridge Regression ensemble using the MATLAB/OCTAVE library for multi-class classification in the MLC Toolbox [21];
- LIFT, multi-label learning with Label specIfic FeaTures) [22];
- Group Preserving Label Embedding (GR) [23];

- LSTM;
- Tab (label for the FastAI Tabular model) [15];
- hML [16].

In the cell Tab-FRAKEL, the reported value was obtained by transforming the original 1024 bit feature vector into 64 int16 features since the original descriptor gained very low performance (0.3165). To avoid overfitting, default parameters were used for the classifiers.

Table 1. Absolute true rates achieved by the classifiers trained on the three descriptors.

| Absolute True | DDI | NRAKEL | FRAKEL |
|---|---|---|---|
| RR | 0.5127 | 0.6062 | 0.5006 |
| LIFT | 0.6111 | 0.5282 | 0.3579 |
| GR | 0.4991 | 0.6093 | 0.4963 |
| LSTM | **0.6626** | 0.6585 | 0.6330 |
| Tab | 0.6441 | **0.7422** | **0.6760** |
| hML | 0.5710 | 0.6791 | 0.5977 |

Examining the results in Table 1, Tab is the best standalone approach, producing an outstanding 0.7422 absolute true rate using NRAKEL descriptors. Of note as well is LSTM, which produced good results on all three descriptors.

The second experiment, reported in Table 2, considers the following ensembles:

- LS, a stacking method based on the approach described in section 2, where LSTM is used as a feature extractor, and the resulting descriptors are given as input to an hML classifier;
- X+Y, fusion by the average rule between the methods X and Y;
- eLS, an ensemble generated by randomly perturbing features obtained as the fusion of ten LS methods trained using random rearrangements of the input features.

Table 2. Absolute true rates achieved by the ensembles on the three descriptors

| Absolute True | DDI | NRAKEL | FRAKEL |
|---|---|---|---|
| LS | 0.6902 | 0.7092 | 0.6709 |
| eLS | 0.6995 | 0.7177 | 0.6853 |
| LSTM+hML | 0.6647 | 0.7371 | 0.6716 |
| eLS+LSTM+hML | 0.6915 | 0.7358 | 0.6894 |
| eLS+LSTM+hML +Tab | 0.6928 | 0.7538 | 0.7072 |

Results reported in Table 2 show a strong performance improvement for descriptors trained on LS (a single hML classifier trained with LSTM features) compared to eLS (an ensemble of ten LS classifiers). The best performance is obtained by the ensemble eLS+LSTM+hML+Tab, which is the fusion of methods with the greatest diversity, compared to the others. This ensemble produces the highest performance in this classification problem, outperforming all the standalone approaches for each of the three descriptors.

In the third experiment (see Table 3), fusion at the feature level is tested. The starting descriptor is the concatenation of two or three sets of features for the Tab approach, while for other classifiers, the combination is the average rule applied to each of them (e.g., LSTM trained on DDI is combined by average rule with LSTM trained on NRAKEL).

When a cell in Table 3 spans more than one column, that indicates that the related classifier is trained using more features, and, for each feature, a different classifier is trained with results fused using the average rule.

Table 3. Combinations of descriptors (absolute true rates) achieved by the ensembles using combinations of features.

| Absolute True | DDI | NRAKEL | FRAKEL |
|---|---|---|---|
| Tab |  | 0.7667 | --- |
| Tab |  | 0.7734 |  |
| eLS+ LSTM+hML |  | 0.7577 | --- |
| eLS+ LSTM+hML |  | 0.7762 |  |
| eLS+LSTM+hML +2×Tab |  | 0.7919 |  |
| eLS+LSTM+hML +3×Tab |  | **0.8009** |  |
| LS+L_M+Tab |  | 0.7812 |  |

The results reported in Table 3 show the usefulness of the ensemble: all the approaches that contain Tab outperform the Fast.AI research group, which has achieved the highest classification score to date.

Finally, in Table 4, we report a comparison of our proposed method with the literature. Clearly, our ensemble strongly outperforms the other approaches. Compare the performance difference of the original papers on NRAKEL [24] and FRAKEL [18] and the classifiers tested in this work. The main reason for this difference is that the classifiers were not optimized here since we are using a single dataset. Our concern in this regard is to avoid any risk of overfitting by running the approaches using default values.

Table 4. Comparison of the best ensemble here with the best reported in the literature.

| Method | Aiming | Coverage | Accuracy | Absolute True | Absolute False |
|---|---|---|---|---|---|
| eLS+ LSTM+hML +3*Tab | **0.9139** | **0.8432** | **0.8338** | **0.8009** | 0.0131 |
| Chen et al. [7] | 0.5076 | 0.7579 | 0.4938 | 0.1383 | 0.0883 |
| EnsANET_LR [12] | 0.7536 | 0.8249 | 0.7512 | 0.6668 | 0.0262 |
| EnsLIFT [10] | 0.7818 | 0.7577 | 0.7121 | 0.6330 | 0.0285 |
| iATC-mISF [8] | 0.6783 | 0.6710 | 0.6641 | 0.6098 | 0.0585 |
| iATC-mHYb [9] | 0.7191 | 0.7146 | 0.7132 | 0.6675 | 0.0243 |
| iATC_Deep-mISF [25] | 0.7470 | 0.7391 | 0.7157 | 0.6701 | **0.0000** |
| NRAKEL [24] | 0.7888 | 0.7936 | 0.7786 | 0.7593 | 0.0363 |
| FRAKEL [18] | 0.7851 | 0.7840 | 0.7721 | 0.7511 | 0.0370 |
| NLSP [26] | 0.8135 | 0.7950 | 0.7828 | 0.7497 | 0.0343 |
| FUS3 [13] | 0.8755 | 0.6973 | 0.7346 | 0.6871 | 0.0238 |

**6 Conclusion**

Since ATC classification is a difficult multi-label problem, the goal of this study was to improve performance by generating ensembles trained on three different feature vectors. The original input vectors were fed into a BiLSTM, which functioned (with modification) not only as a multi-label classifier but also as a feature extractor, with features taken from the output layer.

Two other classifiers aside from LSTM were evaluated: one based on Multiple Linear Regression and another a deep learning technique for tabular/structured data based on the creation of some embedding layers for categorical variables. To boost the performance of these classifiers, they were trained on the feature sets with results fused via average rule. Comparisons of the best ensembles were made with the standalone classifiers and other notable systems. Results show that the top-performing ensemble constructed by the method proposed here obtained superior results for ATC classification using five performance indicators.

Future work will explore the performance of different LSTM and CNN topologies combined using many activation functions. The fusion of other deep learning topologies for extracting features will also be the focus of an investigation.

**Acknowledgment:** This study ran experiments on a TitanX GPU donated by the NVIDIA GPU Grant Program.

**Author Contributions:** Conceptualization, L.N.; methodology, L.N.; software L.N.; writing—original draft preparation, S.B., A.L., and L.N.; writing—review and editing, S.B., A.L., and L.N. All authors have read and agreed to the published version of the manuscript.